\documentclass[conference]{IEEEtran}

\renewcommand{\baselinestretch}{0.999}

\IEEEoverridecommandlockouts
\usepackage{cite}
\usepackage{amsmath,amssymb,amsfonts}
\usepackage{algorithmic}
\usepackage{graphicx}
\usepackage{textcomp}
\usepackage{xcolor}
\usepackage{booktabs}
\usepackage{multirow}
\usepackage{gensymb}
\usepackage{subcaption}

\def\BibTeX{{\rm B\kern-.05em{\sc i\kern-.025em b}\kern-.08em
    T\kern-.1667em\lower.7ex\hbox{E}\kern-.125emX}}
\begin{document}

\title{Performance Analysis of LoRa and LR-FHSS in 3GPP LEO Satellite and HAPS Channel Scenarios for Remote Area IoT Coverage}

\author{\IEEEauthorblockN{Jean Michel de Souza Sant'Ana, Nurul Huda Mahmood and Vinicius Pozzobon Borin}%
\IEEEauthorblockA{\textit{Centre for Wireless Communications, University of Oulu}, Oulu, Finland\\
\{jean.desouzasantana, nurulhuda.mahmood, vinicius.borin\}@oulu.fi}}

\maketitle

\begin{abstract}
In this paper, we analyze and compare the performance of LoRa and LR-FHSS in a remote area Internet-of-Things (IoT) connectivity scenario using a low-earth orbit (LEO) satellites and high-altitude platform station (HAPS)-mounted gateway. We conducted a link budget analysis to evaluate communication stability to assess the ability of the technologies to support the different distances and behaviors using a 3GPP channel model for non-terrestrial networks. Several parameters are considered, such as altitude, minimum elevation angle as well as the impact of different types of loss, such as shadow and fast fading, ionosphere scintillation loss, for example. Our findings highlight the performance gap between LR-FHSS and LoRa in terms of both connectivity. We also highlight how lower HAPS altitudes can benefit certain LoRa settings compared to LEO satellites. We show that HAPS can enable the use of lower LoRa spreading factors, which improves scalability and energy efficiency, while LR-FHSS and a higher spreading factor LoRa remain the solution for LEO satellites.
\end{abstract}


\section{Introduction}

The Internet-of-Things (IoT) connectivity in remote areas is a major challenge for future networks. The integration of terrestrial networks with non-terrestrial networks (NTN) has been proposed as a potential solution to bring access to these regions~\cite{Centenaro:CST:2021}. Most often, devices transmit directly or indirectly (through an aggregator) to a low-earth orbit (LEO) satellite~\cite{Fraire:ComMag:2022}. One of the most discussed communication technologies in this scenario is Long Range (LoRa), the physical layer (PHY) of the LoRa Wide Area Network (LoRaWAN)~\cite{Asad:OJCS:2024}. Most recently, LoRaWAN released a new specially designed PHY for satellite communication with higher capacity called the Long-Range Frequency Hopping Spread Spectrum (LR-FHSS)~\cite{Boquet:ComMag:2021}.

Some limitations arise with the use of LEO satellites for IoT. For example, due to their non-stationary nature, it is hard to guarantee the coverage of a specific area at a given time~\cite{Baltaci:CST:2021}. The difficulties in complying with local legislation, for example, the ISM bands in use and the transmission power for LoRaWAN, pose a challenge for satellites that intend to cover different regions around the globe. Thus, High-Altitude Platform Stations (HAPS) have emerged as a solution to address these limitations~\cite{Kurt:CST:2021}. They provide better communication links with lower path loss due to lower altitudes (8 to 50 km) and improved line-of-sight (LOS) compared to LEO satellites, which can be crucial for devices with limited power budget~\cite{Abbasi:WC:2024}. Moreover, their altitude still allows them to cover large areas with strong LOS coverage. Finally, HAPS can be strategically deployed to provide coverage in specific regions, such as in response to hazards and emergencies.

Several studies address the use of HAPS in IoT scenarios. The authors in~\cite{Qin:TVT:2021} consider a heterogeneous IoT scenario and tackle it with a mixed strategy of Low Altitude Platform Station (LAPS) for more stringent applications and HAPS for broader coverage. 
The authors in~\cite{Zhang:OJCS:2023} discuss the coverage of rural areas with heterogeneous aerial networks, using HAPS to serve as the backhaul for unmanned aerial vehicles (UAVs) and terrestrial base-stations. The backhaul link is optimized such that the freshness of the information is considered. Only a few studies have been conducted on LoRa connectivity issues in HAPS scenarios. For instance, the authors in~\cite{Almarhabi:ICIAS:2021} evaluated the performance of the LoRa link with HAPS considering path loss and different altitudes. Similarly, the work in~\cite{Giambene:SoftCOM:2022} considers LoRa to HAPS in remote areas, also including path loss and different altitudes, but using a simplified collision scheme to account for the scalability of the network. To the best of our knowledge, no studies have included LR-FHSS in a HAPS scenario.

\subsection{Contributions}
In this work, we analyze and compare the feasibility and performance of the two LoRaWAN uplink PHY techniques, LoRa and LR-FHSS, considering a LEO and HAPS IoT scenario. We check the performance as a function of the distance between the ground devices and the gateway for different elevation angles. We consider different altitudes and their impact. Unlike~\cite{Almarhabi:ICIAS:2021,Giambene:SoftCOM:2022}, we consider a shadowing plus fading model~\cite{3GPP} that considers the elevation angle and also introduces the LR-FHSS analysis. Unmanned Aerial Vehicles (UAVs) were not included in the analysis because, due to their short communication distances and strong line-of-sight conditions, the adopted model resulted in an outage probability equal to zero. Under these circumstances, their inclusion would not provide meaningful comparative insights, since the metric could not capture performance differences.

\subsection{Organization}
The remainder of this paper is organized as follows. Section~\ref{sec:2} overviews LoRa and LR-FHSS while Section~\ref{sec:3} presents the system model, the communication assumptions and condition for unsuccessful communication. Section~\ref{sec:4} evaluates some numerical results, while Section~\ref{sec:5} concludes the paper. 

\section{LoRa and LR-FHSS} \label{sec:2}

In this section, we provide details about the two physical layer uplink technologies of LoRaWAN: LoRa and LR-FHSS. Although LoRa can also be used for downlink, in this paper we consider only the uplink, i.e., communication from a device to a HAPS-mounted gateway. 

LoRa consists of a chirp spread spectrum (CSS) modulation technique, where devices can choose one of six spreading factors (SF) (from SF7 to SF12) prior to transmission, with a fixed bandwidth, usually 125 or 250 kHz, in LoRaWAN. The smaller the SF, the higher the transmission rate at the cost of reduced noise resilience. This results in shorter transmission time, or time-on-air, albeit at poorer radio sensitivity. Recent works on LoRa in NTN scenarios have shown that higher SF (usually from SF10 to SF12) are suitable for the long distances from a device to the non-terrestrial gateway, but may also suffer more from Doppler effects in satellite systems~\cite{Asad:OJCS:2024}. Conversely, smaller SFs provide a shorter communication range but greater Doppler immunity~\cite{Zadorozhny:ACS:2022}.

\begin{table}[tb]
\vspace{0.03in}
\caption{LoRa and LR-FHSS parameters}
\label{tab:lora_lrfhss}
\centering
\small
\begin{tabular}{@{}cccc@{}}
\toprule
\multicolumn{2}{c}{\textbf{PHY}}          & \textbf{\begin{tabular}[c]{@{}c@{}}Sensitivity\\ {[}dBm{]}\end{tabular}} & \textbf{\begin{tabular}[c]{@{}c@{}}Bandwidth\\ {[}kHz{]}\end{tabular}} \\ \midrule
\multirow{6}{*}{\textbf{LoRa}}    & SF7                                                                     & -123                                                                     & 125                                                                    \\
                                  & SF8                                                                     & -126                                                                     & 125                                                                    \\
                                  & SF9                                                                    & -129                                                                     & 125                                                                    \\
                                  & SF10                                                                   & -132                                                                     & 125                                                                    \\
                                  & SF11                                                                   & -134.5                                                                   & 125                                                                    \\
                                  & SF12                                                                   & -137                                                                     & 125                                                                    \\ \midrule
\multirow{2}{*}{\textbf{LR-FHSS}} & DR8                                                                   & -137                                                                     & 137                                                                    \\
                                  & DR9                                                                   & -137                                                                     & 137                                                                    \\ \bottomrule
\end{tabular}
\end{table}

LR-FHSS employs frequency hopping spread spectrum with 2-Gaussian minimum-shift keying (GMSK)  modulation. During transmission, the signal hops across multiple 488 Hz bandwidth frequencies, known as physical channels. The total bandwidth depends on regional regulations. An LR-FHSS packet consists of a header and a payload. To enhance reliability, the header can be transmitted two or three times. Moreover, the payload is encoded with a Viterbi convolutional code with rate $1/3$ or $2/3$ and split into 50-bit chunks. LR-FHSS DR8 uses a coding rate of $1/3$ and three header copies, while DR9 adopts a coding rate of $1/2$ with two header copies. Finally, each header copy and payload fragments is transmitted using a different physical channel, following a pseudorandom sequence. The sequence plus all the replications and coding parameters is contained in the header. Thus, for a successful transmission, at least one header must be successfully decoded. In addition to that, at least 1/3 or 2/3 of the payload fragments, depending on the coding rate, need to be received to reconstruct the original payload. Unlike LoRa, LR-FHSS lacks a parameter that directly improves the communication range. However, like LoRa, certain parameters increase the packet duration. This enhances transmission resilience by increasing the number of header replicas or decreasing the coding rate, though at the cost of longer time-on-air and collision probabilities. 

The general parameters of LoRa and LR-FHSS can be seen in Table~\ref{tab:lora_lrfhss}. Overall, LR-FHSS offers a slightly higher link budget, comparable to LoRa SF12, and provides additional diversity through header replication and payload coding with frequency hopping, making LR-FHSS transmissions longer, on average, than LoRa. This behavior implies that LR-FHSS might consume more energy for each transmission than LoRa. Although not explored in this paper, this is also an important factor to consider. For a more detailed comparison of energy consumption between LoRa and LR-FHSS, we refer the reader to the work in~\cite{sanchez-vital:sensors:2025}.

\section{System Model}\label{sec:3}

\begin{figure}[tb]
    \centering
    \includegraphics[width=0.9\linewidth]{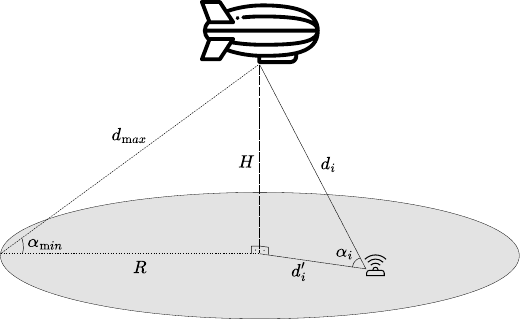}
    \caption{System model.}
    \label{fig:system}
\end{figure}

We follow the system model in \cite{3GPP}, which consists of a gateway-mounted HAPS or LEO satellite at height $H$, over an area, which footprint (the area where communication is possible) covers a circular region of radius $R$ with $N$ devices uniformly deployed at random. Assuming the spherical Earth model, the distance between the gateway and a given device $i$ (slant range) $d_i$ is given as
\begin{align}
    d_i = \sqrt{R_E^2\sin^2\alpha_i + H^2 + 2HR_E} - R_E\sin\alpha_i,
\end{align}
where $\alpha_i$ is the elevation angle between the gateway and device $i$ as
\begin{align}
\alpha_i =
\arctan\left(
\frac{
\cos\left(\dfrac{d_i^{'}}{R_E}\right)
-
\dfrac{R_E}{R_E+H}
}{
\sin\left(\dfrac{d_i^{'}}{R_E}\right)
}
\right),
\end{align}
where $d_i^{'}$ is the distance from the device to the gateway projection on earth (nadir point), as shown in \figurename~\ref{fig:system}.
Given a minimum elevation angle for successful communication $\alpha_{\min{}}$, which occurs at the edge of the footprint coverage area, we can establish a relationship between the footprint radius $R$ and the gateway height as
\begin{align}
    R = R_E\cos^{-1}\left(\frac{R_E}{R_E+H}\cos\alpha_{\min}\right)-\alpha_{\min}. \label{eqn:radius}
\end{align}

We assume fixed transmit power $p_t$, bandwidth\footnote{There is a small difference in the LoRa and LR-FHSS bandwidth. For the ease of comparison, we assume they are equivalent, as the difference is less than $10\%$.} $B$ and carrier frequency $f$. Following the 3GPP channel model for non-terrestrial networks~\cite{3GPP}, we model the received power of a single transmission at the gateway from device $i$ in dB as
\begin{align}
    p_{rx}^{i} &=p_t + G_t + G_r - FSPL(d_i)- SF(\alpha_i) \\ \nonumber
    & - CL(\alpha_i) - SL - 10\log(|FFL(\alpha_i)|^2),
\end{align}
where $G_t$ and $G_r$ are the transmitter and receiver antenna gains, $FSPL(d)$ is the free-space path loss component, $SF$ is the shadowing fading, $CL$ is the clutter loss, $SL$ is the scintillation loss and $FFL$ is the fast fading loss component.
The free space path loss given as
\begin{align}
    FSPL(d) = 32.45 + 20\log (f_c) + 20\log (d)\label{eqn:pathloss}
\end{align}
where $f_c$ is the carrier frequency in GHz.

In this work, we utilize values equivalent to S-band, as it is the closest to the ISM band which is shown in the 3GPP documents. Also, to account for remote areas, we utilize values from the scenario called ``rural scenarios''. Thus, we summarize the parameters for the following expressions in Table~\ref{tab:fading_param}. Following~\cite{3GPP}, the parameters are taken from the nearest reference elevation angle.

\begin{table}
\centering
\caption{Channel model parameters}
\begin{tabular}{c|cccc} 
\toprule
$\alpha_i$ & \begin{tabular}[c]{@{}c@{}}LOS\\probability\end{tabular} & $CL$ & $\mu_K$ & $\sigma_K$  \\
\hline
10°       & 78.2\%                                                   & 19.52  & 25.43                     & 7.04                          \\
20°       & 86.9\%                                                   & 18.17  & 12.72                     & 7.47                          \\
30°       & 91.9\%                                                   & 18.42  & 8.40                      & 7.18                          \\
40°       & 92.9\%                                                   & 18.28  & 6.52                      & 6.88                          \\
50°       & 93.5\%                                                   & 18.63  & 5.24                      & 5.28                          \\
60°       & 94.0\%                                                   & 17.68  & 4.57                      & 4.92                          \\
70°       & 94.9\%                                                   & 16.50  & 4.02                      & 3.40                          \\
80°       & 95.2\%                                                   & 16.30  & 3.70                      & 2.22                          \\
90°       & 99.8\%                                                   & 16.30  & 3.62                      & 2.28                          \\
\bottomrule
\end{tabular}
\label{tab:fading_param}
\end{table}

The shadowing fading is modeled as a log-normal distribution, which in dB is expressed as a zero-mean normal distribution with standard distribution $\sigma^2_{SF}$ given as
\begin{align}
    \sigma^2_{\mathrm{SF}}=
    \begin{cases}
        4~\text{dB}, & \text{if LOS}, \\
        8~\text{dB}, & \text{if NLOS}.
    \end{cases}
\end{align}
Note that we assign a transmission to LOS following the probabilities in Table~\ref{tab:fading_param}.

We consider ionospheric scintillation loss only, following \cite{3GPP} as the carrier frequency is below 6 GHz. In addition, ionospheric scintillation is only relevant for LEO satellites, since HAPS operate at altitudes lower than the ionosphere (usually above 60 km). For that, we consider a $S_4$ level of 0.2, which represents a weak scintillation scenario. We derive the peak fluctuation amplitude, scale it to the desired carrier frequency and generate the signal loss using (6.6-14), (6.6-11) and (6.6-13) in \cite{3GPP}, resulting in
\begin{align}
    SL = 
    \begin{cases}
        9.60~\text{dB}, & \text{if LEO}, \\
        0~\text{dB}, & \text{if HAPS}.
    \end{cases}
\end{align}

Finally, the fast fading loss is modeled as
\begin{align}
    FFL \sim
    \begin{cases}
        \text{Rice}(\mu_K, \sigma_K), & \text{if LOS}, \\
        \text{Rayleigh}, & \text{if NLOS}.
    \end{cases}
\end{align}

\subsection{Condition for unsuccessful communication}

We analyze the probability that the received signal is below the sensitivity threshold to successfully decode the transmission. This relates to the technology capacity to achieve long distances given its parameter settings. It is given as
\begin{align}
    C_1 = p_{rx}^{i} < \gamma,
\end{align}
where $\gamma$ is the technology-dependent sensitivity threshold, which changes according to LoRa SF for LoRa, while it is the same for any configuration of LR-FHSS. This condition is also affected by the gateway altitude, the relative positioning between device and gateway, such as the elevation angle.

\section{Numerical Results} \label{sec:4}

This section presents numerical results obtained with Monte Carlo simulations. We consider the parameters for LoRa and LR-FHSS according to Table~\ref{tab:lora_lrfhss}. Moreover, unless otherwise stated, we consider a transmission power of $p_t=14$~dBm, transmitter and receiver antenna gains of $G_t=0$~dBi and $G_r=5$~dBi, respectively, a minimum elevation angle of $\alpha_{\min{}}=20\degree$. The Monte Carlo simulation results are based on an average of $10^8$ simulation points.

\begin{figure*}[]
    \centering
    \begin{subfigure}{\columnwidth}
    \includegraphics[width=0.95\linewidth]{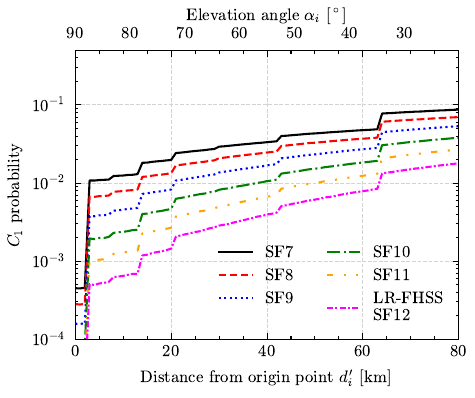}
    \caption{$H=30$~km}
    \label{fig:success_heights_30}
    \end{subfigure}
    \begin{subfigure}{\columnwidth}
    \includegraphics[width=0.95\linewidth]{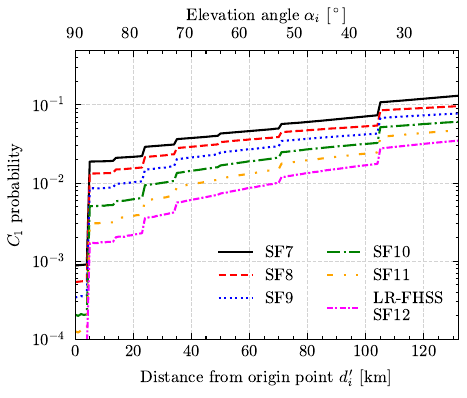}
    \caption{$H=50$~km}
    \label{fig:success_heights_50}
    \end{subfigure}
    \caption{Condition 1 ($C_1$) probability versus the distance from the origin point $d_i'$ and elevation angle $\alpha_i$ for HAPS with different heights $H$ and LoRa spreading factors (SF) and LR-FHSS.}
    \label{fig:success_heights}
\end{figure*}

\begin{figure*}[]
    \centering
    \begin{subfigure}{\columnwidth}
    \includegraphics[width=0.95\linewidth]{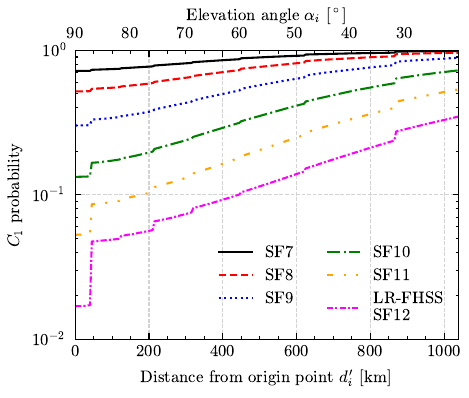}
    \caption{$H=500$~km}
    \label{fig:success_heights_500_sat}
    \end{subfigure}
    \begin{subfigure}{\columnwidth}
    \includegraphics[width=0.95\linewidth]{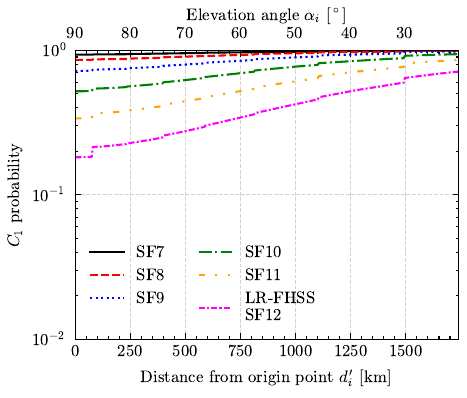}
    \caption{$H=1000$~km}
    \label{fig:success_heights_1000_sat}
    \end{subfigure}
    \caption{Condition 1 ($C_1$) probability versus the distance from the origin point $d_i'$ and elevation angle $\alpha_i$ for LEO satellite with different heights $H$ and LoRa spreading factors (SF) and LR-FHSS.}
    \vspace{-12pt}
    \label{fig:success_heights_sat}
\end{figure*}

\figurename~\ref{fig:success_heights} and \figurename~\ref{fig:success_heights_sat} depicts the $C_1$ condition probability as a function of the distance from the origin point $d_i$ for HAPS and LEO satellite scenarios, respectively. It is also a function of the elevation angle $\alpha_i$, since $\alpha_i$ and $d_i$ are directly related. The maximum distance is set to correspond to the minimum elevation angle $\alpha_{\min}=20$. The discontinuities presented matches the elevation angle changes from Table~\ref{tab:fading_param}, which changes the LOS probability and the fading parameters.

In~\figurename~\ref{fig:success_heights} we can see the impact of the SFs in the $C_1$ probability. Also, it is possible to see the degradation higher altitudes bring, although increasing the coverage footprint by over 50\% changing from 30~km to 50~km. Although higher SFs and LR-FHSS present the best coverage quality (smaller $C_1$), it is possible to state that the other spreading factors, such as SF7 and SF8 can achieve outage probabilities smaller than $10^{-1}$ for the most part of the network. 
In contrast, \figurename~\ref{fig:success_heights_sat} depicts the LEO scenario with much worse performance. This is due the higher altitudes, and thus, longer distances, and some losses atributted to it, like the scintillation loss, which is not present in the HAPS scenario. The results match some experiments, where only higher spreading factors being able to guarantee communication with the satellite. Moreover, we can see that using lower orbits in the LEO layer might be important, as with $H=1000$~km, the outage remains above $10^{-1}$ even for SF12 and LR-FHSS.

\begin{figure}
    \centering
    \includegraphics[width=0.95\linewidth]{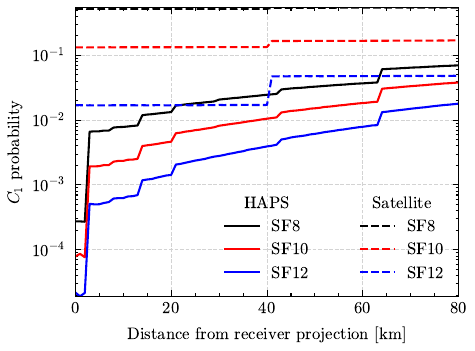}
    \caption{Condition 1 ($C_1$) probability for HAPS with $H=30$ km and LEO satellite with $H=500$ km as a function of the distance from origin point $d_i^{'}$.}
    \label{fig:HAPS_LEO}
\end{figure}

Different from HAPS, that are mostly hovering over the same place, LEO satellites are in constant movement. Thus, in the aforementioned HAPS figures, we can understand the x-axis as a spatial axis, in which for better transmission performance, a device would need to get closer to the nadir point. However, in the LEO figure, we can understand the x-axis as time axis, where devices should choose the right moment to transmit, so that the LEO satellite is in a better position for transmission. Thus, in \figurename~\ref{fig:HAPS_LEO}, we compare HAPS and LEO satellite performance. Here we are assuming if the same devices transmitted to either the HAPS or LEO satellite, thus the network radius was set to the HAPS scenarios similar to \figurename~\ref{fig:success_heights}. Here we can see that, even if the device in a LEO scenario chooses properly when to transmit, the HAPS scenario still outperforms LEO satellite scenario in all cases. Again, we can see that lower SF do not perform in the LEO satellite scenario, performing well only with SF12.

\begin{figure}
    \centering
    \includegraphics[width=0.95\linewidth]{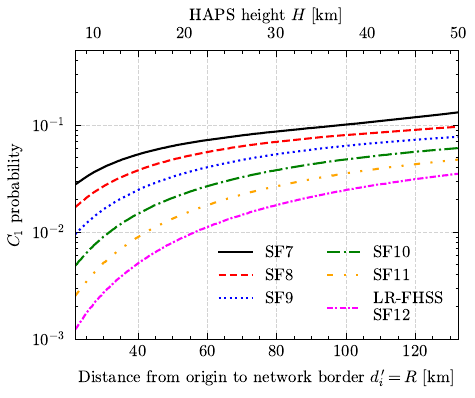}
    \caption{Condition 1 ($C_1$) probability at the network border, when $\alpha=\alpha_{\min{}} = 20\degree$}
    \label{fig:success_fixed_angle}
\end{figure}

\figurename~\ref{fig:success_fixed_angle} depicts the $C_1$ probability at the network border (when $\alpha_i = \alpha_{\min}$) for different HAPS heights. Therefore, it means also that the distance from origin to the device $d_i$ increases, as the network radius $R$ increases with $H$, from \eqref{eqn:radius}. This is an important analysis, as we can see the worst-case performance for all different HAPS heights, especially when one wants to guarantee some minimum performance requirements. These curves show that a lower altitude can ensure better reliability levels for the smaller spreading factors. For example, in the case of SF7, a performance below 0.1 can be observed with a 30~km high HAPS. Also, for performances below $10^{-2}$, SF12 reaches at around 20~km high. However, this comes at the cost of a smaller coverage footprint, which means a reduced coverage area.

\section{Conclusions}\label{sec:5}

In this work, we compare LoRa and LR-FHSS for remote area coverage using LEO satellites and HAPS-mounted gateways. Additionally, we analyze how key network parameters, such as altitude and coverage area, impact network performance while using 3GPP channel models.
We show that lower spreading factors can be employed in HAPS scenarios. Also, since LR-FHSS brings a complexity and energy cost, LoRa may be better suited for HAPS communication, while in LEO satellite scenarios, LR-FHSS appears as a better option.
Finally, we expect these parameters to be further explored in future research and deployments to enhance reliability and coverage.

\section*{Acknowledgements}
{\small This work or its authors have been partially supported in Finland by the Research Council of Finland 6G Flagship Programme (Grant Number: 369116), and by the European Union through the Interreg Aurora project ENSURE-6G (Grant Number: 20361812).}

\bibliographystyle{IEEEtran}
\bibliography{references}

\end{document}